\documentclass{acmsiggraph}
\newcommand{\blfootnote}[1]{%
  \begingroup
  \renewcommand{\thefootnote}{}%
  \footnotetext{#1}%
  \addtocounter{footnote}{-1}%
  \endgroup
}
\title{Substepping the Material Point Method}
\author{Chenfanfu Jiang$^*$
 \\ University of California, Los Angeles}
\pdfauthor{J}
\usepackage{setspace,subfigure,afterpage,fix-cm,multirow}
\usepackage{algorithm}
\usepackage[noend]{algpseudocode}
\usepackage{booktabs}
\usepackage{multirow}
\begin{document}
\maketitle

\begin{abstract}
Many Material Point Method implementations favor explicit time integration. However large time steps are often desirable for special reasons -- for example, for partitioned coupling with another large-step solver, or for imposing constraints, projections, or multiphysics solves. We present a simple, plug‐and‐play algorithm that advances MPM with a large time step using substeps, effectively wrapping an explicit MPM integrator into a pseudo‐implicit one.
\end{abstract}

\section{Introduction}

\blfootnote{$^*$email: chenfanfu.jiang@gmail.com}

Explicit MPM is straightforward and GPU‐friendly but requires small time steps. Implicit MPM permits large steps but requires nonlinear solves with sparsity patterns that change topologically at each step \cite{jiang2016material}. In many practical cases, without advanced preconditioning and agreesive code-level optimization, the speedup gained from larger time steps is outweighed by the cost of the implicit solve, often making a fully implicit formulation not worthwhile; it also entails considerable implementation overhead.

However, sticking with explicit MPM has drawbacks. One case is coupling MPM with a rigid, fluid, or FEM solver that can take large time steps. This leads to mixed implicit-explicit asynchronous coupling and demands custom, complicated algorithmic designs. Another case arises when one wishes to solve additional physics alongside MPM elastodynamics, such as an incompressibility projection or thermodynamics. These problems naturally prefers implicit solvers, and solving them across small steps is costly.

Going from $t^n$ to $t^{n+1} = t^n+\Delta t$, one might try running $S$ explicit MPM substeps with $\delta t = \Delta t/S$, producing grid velocities $v_i^{n+1,1}, v_i^{n+1,2}, \dots, v_i^{n+1,S}$, and then simply taking $v_i^{n+1,S}$ as $v_i^{n+1}$. While such an approach could work for FEM, it fails for MPM because particles move during substeps and grid allocations are recomputed. Thus, unlike the desired $v_i^{n+1}$, which must be defined over $\Omega^n$ (the reference configuration), $v_i^{n+1,S}$ is defined over a different grid associated with the substepped state, whose degrees of freedom generally do not align with those at time $n$.

In this note we present an algorithm that finds a $v_i^*$ defined over $\Omega^n$ such that setting $v_i^{n+1}=v_i^*$ faithfully mimics what occurred over the substeps. Consequently, one obtains a black‐box routine that can be treated as an implicit MPM integrator.

\section{Method}

We formulate the problem as a least‐squares fit of desired particle velocities and velocity gradients at $t^{n+1}$ over $\Omega^n$. We define the
\emph{secant} particle velocity and velocity gradient targets based on particle positions and deformation gradients before and after the substeps:
\begin{equation}
\boxed{
\Delta t v_p^\star \;=\; x_p^{n+1,S}-x_p^n,
\qquad
\Delta t C_p^\star \;=\; F_p^{n+1,S}(F_p^n)^{-1}-I.
}
\label{eq:secant_targets_LS}
\end{equation}
We then seek $v_i^*$ such that advecting particles with $v_i^*$ reproduces these targets. This leads to
\begin{equation}
\min_{\{v_i\}}\;
\underbrace{\sum_p m_p\|v_p(v)-v_p^\star\|^2}_{\text{velocity match}}
\;+\;
\lambda\underbrace{\sum_p m_p\|(\nabla v)_p(v)-C_p^\star\|_F^2}_{\text{gradient match}},
\label{eq:ls_objective}
\end{equation}
where $v_p(v)=\sum_i w_{ip}\,v_i$ and $(\nabla v)_p(v)=\sum_i v_i\,(\nabla w_{ip})^\top$.

Stationarity with respect to a node $k$ gives
\begin{equation}
\sum_p m_p\,w_{kp}\,\big(v_p-v_p^\star\big)
+
\lambda\sum_p m_p\big[(\nabla v)_p-C_p^\star\big](\nabla w_{kp}) \;=\; 0.
\label{eq:normal_eq_compact}
\end{equation}
Expanding $v_p$ and $(\nabla v)_p$ yields a sparse SPD grid system:
\begin{align}
\sum_j
\Big[
\underbrace{\sum_p m_p\,w_{kp}w_{jp}}_{\text{``mass''}}
\;+\;
\lambda\underbrace{\sum_p m_p\,(\nabla w_{kp})\!\cdot\!(\nabla w_{jp})}_{\text{``stiffness''}}
 \Big] v_j \nonumber  \\
=
\sum_p m_p\,w_{kp}\,v_p^\star \;+\; \lambda\sum_p m_p\,C_p^\star(\nabla w_{kp}).
\label{eq:normal_eq_expanded}
\end{align}
Applying \emph{mass lumping}, $\sum_p m_p w_{kp}w_{jp}\approx \delta_{kj} m_k$, dropping the stiffness on the left‐hand side, and using $\nabla w_{kp} \;\approx\; \tfrac{1}{D}\,w_{kp}\,(x_k-x_p)$, where $D=\tfrac14\,\Delta x^2$ for quadratic B‐splines \cite{jiang2015affine,hu2018moving}, we obtain
\[
m_k v_k \;=\; \sum_p m_p w_{kp} v_p^\star
\;+\;\lambda \sum_p m_p \frac{1}{D} w_{kp}\, C_p^\star (x_k-x_p).
\]
Choosing the natural scaling $\lambda=D$ gives the APIC reconstruction in closed form:
\begin{equation}
\boxed{\quad m_k v_k \;=\; \sum_p m_p w_{kp}\Big(v_p^\star + C_p^\star (x_k-x_p)\Big)\quad}
\end{equation}
i.e., the standard APIC transfer with particle quantities $(v_p^\star, C_p^\star)$, followed by division by $m_k$ and boundary/contact projection.

\section{Discussion}

The method works regardless of whether the underlying explicit MPM scheme is based on FLIP/PIC blends, APIC, or MLS‐MPM. At the macro step, the resulting $v_i^{n+1}$ can also be used with any standard transfer scheme back to the particles. This scheme does not magically make explicit solves faster or more stable -- the CFL limits and substep costs still apply -- and it introduces mild dissipation because the macro-step reconstruction is an extra step that filters high-frequency subgrid content and can slightly mismatch substep energy.

\bibliographystyle{acmsiggraph}
\renewcommand{\refname}{References}
\begingroup
\bibliography{main}
\endgroup

\end{document}